\documentclass[%
 reprint,
superscriptaddress,
nofootinbib,
 amsmath,amssymb,
 aps,
]{revtex4-2}

\usepackage{verbatim}
\usepackage{graphicx}
\usepackage{dcolumn}
\usepackage{bm}
\usepackage{txfonts}
\usepackage{color}
\usepackage{CJK}

\begin{document}
\begin{CJK}{UTF8}{mj}

\preprint{APS/123-QED}

\title{Consistency landscape of network communities}
\author{Daekyung Lee (이대경)}
\affiliation{Department of Physics, Sungkyunkwan University, Suwon 16419, Korea}
\author{Sang Hoon Lee (이상훈)}
\email[Corresponding author: ]{lshlj82@gnu.ac.kr}
\affiliation{Department of Liberal Arts, Gyeongsang National University, Jinju 52725, Korea}
\affiliation{Future Convergence Technology Research Institute, Gyeongsang National University, Jinju 52849, Korea}
\author{Beom Jun Kim (김범준)}
\affiliation{Department of Physics, Sungkyunkwan University, Suwon 16419, Korea}
\author{Heetae Kim (김희태)} 
\email[Corresponding author: ]{kimheetae@gmail.com }
\affiliation{Department of Energy Technology, Korea Institute of Energy Technology, Naju 58322, Korea}
\affiliation{Data Science Institute, Faculty of Engineering, Universidad del Desarrollo, Santiago 7610658, Chile}

\date{\today}

\begin{abstract}
The concept of community detection has long been used as a key device for handling the mesoscale structures in networks. 
Suitably conducted community detection reveals various embedded informative substructures of network topology. 
However, regarding the practical usage of community detection, it has always been a tricky problem to assign a reasonable community resolution for networks of interest.
Because of the absence of the unanimously accepted criterion, most of the previous studies utilized rather \emph{ad hoc} heuristics to decide the community resolution. 
In this work, we harness the concept of consistency in community structures of networks to provide the overall community resolution landscape of networks, which we eventually take to quantify the reliability of detected communities for a given resolution parameter. 
More precisely, we exploit the ambiguity in the results of stochastic detection algorithms and suggest a method that denotes the relative validity of community structures in regard to their stability of global and local inconsistency measures using multiple detection processes. 
Applying our framework to synthetic and real networks, we confirm that it effectively displays insightful fundamental aspects of community structures.
\end{abstract}

\maketitle

\section{\label{sec:intro}Introduction}

Understanding the structure of large networks has been a challenging task across a variety of fields. 
Community detection, which refers to a process of dividing a large network into smaller groups called communities, is one of the most conventional and effective approaches in network science~\cite{Porter1,graph_detection}. 
In the general framework of community detection, several candidate mesoscale structures compete to optimize a certain objective function to better describe the substructures of the target network~\cite{finding,finding_large}.
Thanks to its widespread usage and applicability, researchers have proposed distinct detection algorithms to show the different aspect of network structure~\cite{louvain,fast,spectral,EO,GA,ensemble,GenLouvain,Leiden}. 

Despite its extensive practicality, however, the basic framework of community detection often encounters a substantial ambiguity. 
Since a network generally contains multiple mesoscale structures of various sizes, deciding appropriate resolutions or equivalently typical scales of communities is a practically nontrivial issue. 
Many detection algorithms assign a free parameter to handle the community resolution~\cite{resolution}, but the process of adjusting the parameter usually involves a rather \emph{ad hoc} heuristic specific to networks of interest.
In addition, a network can always harbor communities existing across multiple relevant resolutions. 
In order to observe the full-scale resolution landscape, we need to define a quantitative measure to compare how well the community structures at each resolution describe the target network.

As a way to handle this vagueness, we focus on a fundamental aspect in community detection. 
In the typical process of community detection, it is well known that a lot of existing algorithms adopt the stochastic method to find a suitable community structure of the target network~\cite{consistency}.
In other words, because of the computational complexity of community detection, exploration of the proper mesoscale structure is usually achieved by \emph{local} optimization algorithms that frequently exploit stochastic processes, e.g., greedy algorithms~\cite{louvain}, extremal optimizations~\cite{EO}, genetic algorithms~\cite{GA}, etc. 
In general, the results of those detection algorithms are different for each realization~\cite{ensemble,CoI_PRE}. 
Naturally, questions arose to ``solve'' this problem of ambiguity, e.g., via the comparison between detected community structures~\cite{ensemble,CC1,CC2,EC}. 
Throughout those works, such \emph{inconsistency} in community detection results has been regarded as an inevitable error derived due to the limitation of computational resources. 

Recent studies~\cite{CoI_PRE,Versatility,MPS}, however, have shown that those properties can be exploited as an additional source of genuine information about local community structure. 
For instance, if a node belongs to communities with different groups of nodes in multiple realizations,\footnote{As network communities are usually identified only in terms of their constituent nodes, consistency in communities only concerns the identities of membership nodes regardless of their community indices, which are usually meaningless.} it is considered to play an inconsistent role in the network of interest~\cite{CoI_PRE}.
We notice that the approach can be extended to measure the validity of community structure itself. If a result of a given community detection algorithm is inconsistent or different for each trial, it is likely that the result is less reliable than the case of consistently detected community structures across different trials. In other words, a common community structure found in multiple realizations of stochastic algorithms describes a network with an appropriate scale of reliable communities. Therefore, we propose to use the statistical consistency of community structure as an effective criterion to compare the several candidate structures in networks. By observing the overall \emph{consistency} landscape of network communities as a function of resolution, we identify the most reliable ranges of community resolution and their corresponding community structure.

Systematizing such procedures, in this paper, we propose a framework of using the global and local consistency of community structures to examine the community landscape of networks. Using the statistical variability of community detection results, we construct a series of measures that assign the (in)consistency of community ensembles. Our extensive analysis on model and real networks shows that our method successfully detects diverse forms of mesoscale substructures of the networks, which elucidate the community landscape for given networks in an unprecedented way.
Previously, some studies covered related issues. For instance, in Ref.~\cite{MPS}, the instability of community structure is quantified as the concept of mean partition similarity between different realizations. Based on that similarity, they suggested the relative credibility of each resolution point as their statistical significance with respect to that of the randomized pairs. Although the methodology precedes ours and can be useful to characterize the community structures in various scales, it has recently been shown that their Rand-index-based similarity measure itself faces some issues in terms of distinguishability in the comparison of community configurations~\cite{EC}. Our approach, in contrast, takes the recently developed similarity measure~\cite{EC} and provides a more systematic toolkit for the comprehensive analysis on community structures with a step-by-step instruction set, which will be detailed from now on.

\section{\label{sec:method}Measures of community inconsistency}

\begin{figure*}
\includegraphics[width=\textwidth]{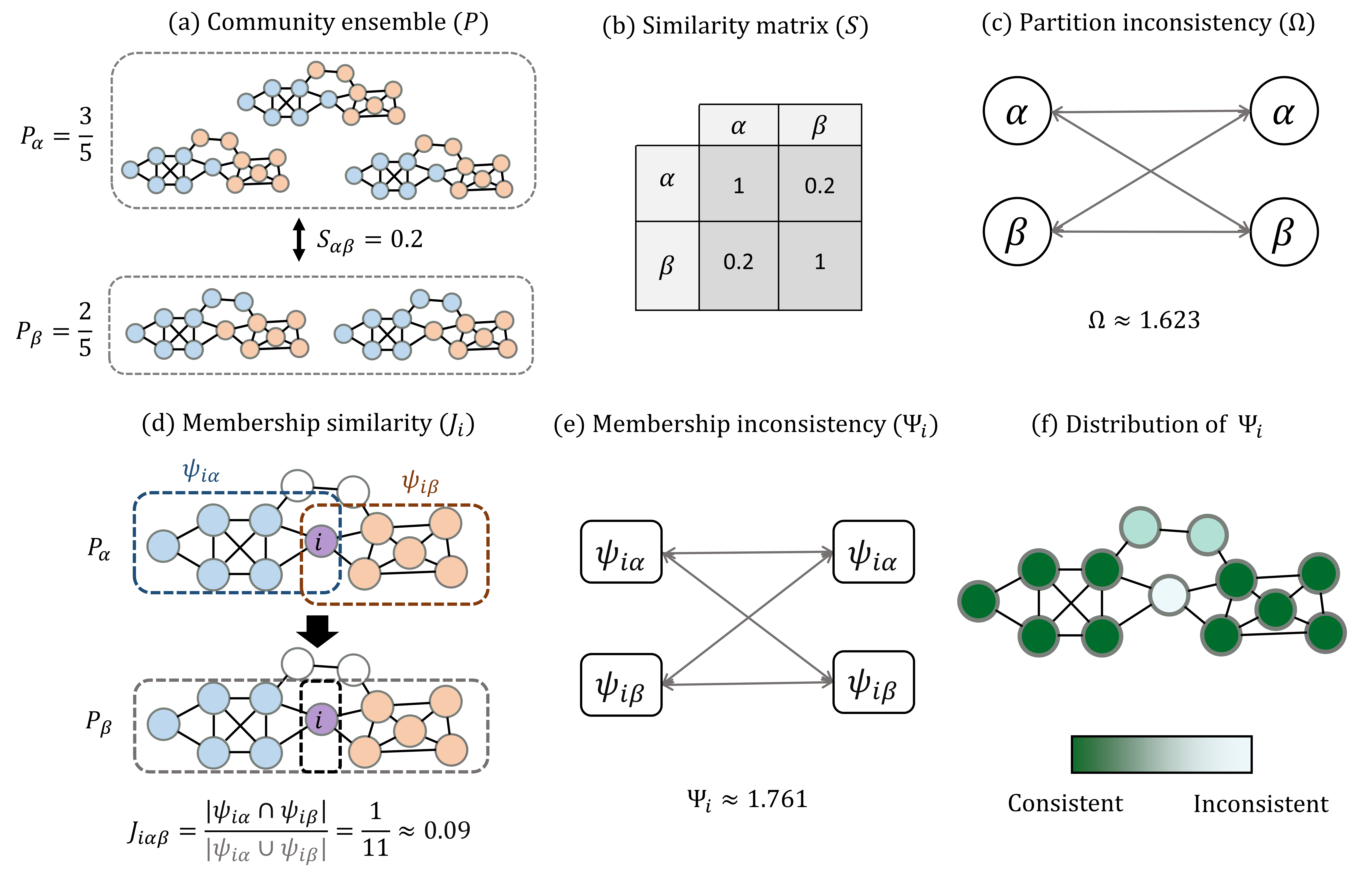}
\caption{Description on the concept of our inconsistency measures. In panel (a), we plot a community ensemble composed of two different community configurations denoted by $\alpha$ and $\beta$, and their relative frequencies of appearance $p_\alpha$ and $p_\beta$, respectively, from five realizations of a stochastic community detection algorithm applied to a sample network. The panel (b) shows the similarity matrix for the configuration pairs described in panel (a), and the partition inconsistency is calculated in panel (c) using the information on the composition and similarity between distinct configurations in the community ensemble. We also describe the definition of membership similarity in panel (d) with the example of node $i$ and two distinct membership sets. In panel (e), the membership inconsistency (MeI) is defined from the membership information of node $i$. The MeI values of the nodes in the sample network are displayed in panel (f).}
\label{fig:concept}
\end{figure*}

We present our approach by explaining stochastic features of community structures in networks. 
In principle, there exists intrinsic ambiguity in the results of community detection, unless the ground-truth community structure is actually given (and in that case, of course, community detection is needless). 
In practice, many community detection algorithms rely on stochastic trials, as the exhaustive enumeration of all of the possible community division is not computationally plausible. 
Such a stochastic nature is usually considered as an obstacle in finding the exact community structure of networks because in principle it can yield different results for each realization of the algorithm. 
However, as our previous work~\cite{CoI_PRE} suggested, the statistical property of results can be an additional source of information, in particular, on the local inconsistency of individual nodes based on the credibility of communities found. 
From now on, we introduce how specifically such inconsistent results of community detection can be a much richer source of information than just a local property.

\subsection{Community ensemble}
\label{sec:community_ensemble}

We describe the situation in terms of statistical ensembles: each realization of a stochastic detection algorithm yields a community detection result or a \emph{configuration}: the list of communities and their constituent nodes, and the relative proportions of distinct community structures compose the probability distribution of each unique configuration of community structures. 
In this work, we take the GenLouvain community detection algorithm~\cite{GenLouvain}, which is a variant of the celebrated Louvain algorithm~\cite{louvain}. 
As a representative stochastic algorithm, GenLouvain can yield different community structures for each realization in principle. 
In the process of our methodology, we collect the results of multiple realizations of the GenLouvain algorithm and take them as a community ensemble. 
In the following, we introduce our theoretical scheme (Fig.~\ref{fig:concept}) and notions to explain the various characteristics of an ensemble of community structures.


\subsection{Global inconsistency}
\label{sec:global_inconsistency}

Let us define several attributes of the statistical community ensemble. 
In order to concisely describe the statistical distribution of each configuration, we first identify a set of distinct community configurations and their relative frequency of appearance in the ensemble. For the clarity in description, we use Latin indices for node indexing and Greek indices (except for $\gamma$, which is conventionally used to denote the community resolution parameter) for community indexing. 
From the total number $m$ of community detection realizations, we denote the number of unique configurations detected by $\mathcal{C}$, and the proportion of each configuration $\alpha \in \{ 1, 2, \cdots, \mathcal{C} \}$ by $p_{\alpha} = m_{\alpha} / m$, where $m_{\alpha}$ is the number of configurations that include $\alpha$.
We also calculate the similarity $S_{\alpha\beta}$ between configurations $\alpha$ and $\beta$, by employing the element-centric similarity (ECS)~\cite{EC}. 
The ECS is a clustering-comparison method designed to satisfy several necessary conditions for mathematically proper comparison between detected communities in networks. 
In the scheme of ECS, the similarity between two configurations is calculated by the comparison for each node pair, which is the systematic projection from cluster affiliations. 
The ECS method first constructs a cluster-induced element graph composed of the edges between all of the node pairs if they belong to the same community in the original network. 

In the cluster-induced element graph of configuration $\alpha$, we assign an attribute of node $i$ with respect to another node $j$, which is the stationary probability distribution $f_{ij}^{\alpha}$ induced by the personalized PageRank algorithm (PPR)~\cite{PPR} applied to node $i$. 
Then the ECS between two community configurations $\alpha$ and $\beta$ is defined as
\begin{equation}\label{eq:EC}
S_{\alpha\beta} = \frac{1}{N}\sum_{i=1}^{N} \left( 1 - \frac{1}{2d}\sum_{j=1}^{N} \left| f_{ij}^{\alpha} - f_{ij}^{\beta}\right| \right) \,,
\end{equation}
where $N$ is the number of nodes and $f_{ij}^{\alpha}$ is node $j$'s relative importance in the stationary state of PPR starting from the node $i$ in configuration $\alpha$, and we use the default value of the damping factor $d = 0.9$ as used in Ref.~\cite{EC}.

The ECS in Eq.~\eqref{eq:EC} is normalized so that it reaches its maximum value of $1$ for the identical configuration pairs and it becomes $0$ when the two configurations are completely independent. With those notations, we define the \emph{partition inconsistency} (PaI) of a community ensemble as
\begin{equation}\label{eq:PaI}
\Omega = \left( \sum_{\alpha=1}^{\mathcal{C}}\sum_{\beta=1}^{\mathcal{C}} p_{\alpha}p_{\beta}S_{\alpha \beta}\right)^{-1} \,,
\end{equation}
which corresponds to the average similarity between all configuration pairs. 
The intuitive meaning of PaI can be revealed by a simple null model case; if all the configurations are independent with each other and emerge with the same uniform probability $p_{\alpha} = 1 / \mathcal{C}$, all of the off-diagonal components of $S_{\alpha \beta}$ vanish and PaI of this ensemble reaches the maximum value of $\mathcal{C}$. 
If every realization of the community detection gives exactly the same community structure, it becomes trivially $\Omega = 1$ (the minimally inconsistent or the maximally consistent case).
Based on the above reasoning, we infer that the PaI value $\Omega$ essentially corresponds to the effective number of independent configurations in the entire ensemble.

\subsection{Local inconsistency}
\label{sec:local_inconsistency}

The same train of logic can also be applied to quantify the inconsistency in the individual node level. 
We first extract the comembership composition of node $i$ in each configuration $\alpha$ denoted by the set
\begin{equation}\label{eq:neighbor}
\psi_{i\alpha} = \{ j | g_{j\alpha} = g_{i\alpha} \} \,,
\end{equation}
where $g_{i\alpha}$ is the community label\footnote{Again, the label itself is not meaningful and only used to formally distinguish the nodes belonging to the same or different communities.} of node $i$ in the $\alpha$th configuration. 
Note that each set contains the index of configuration $\alpha$, so the consistency of node $i$ depends on the stability of $\psi_{i\alpha}$ with respect to $\alpha$. 
Using the same similarity-based approach as PaI, we define the similarity between $\psi_{i\alpha}$ and $\psi_{i\beta}$ (the node $i$'s comembership similarity between configurations $\alpha$ and $\beta$) as 
\begin{equation}\label{eq:J}
J_{i\alpha \beta} = \frac{|\psi_{i\alpha} \cap \psi_{i\beta}|}{|\psi_{i\alpha} \cup \psi_{i\beta}|} \,,
\end{equation}
which is the Jaccard index between the two sets. 
With this similarity index, we define the \emph{membership inconsistency} (MeI) of node $i$ as
\begin{equation}\label{eq:MeI}
\Psi_{i} = 	\left( \sum_{\alpha=1}^{\mathcal{C}}\sum_{\beta=1}^{\mathcal{C}} p(\psi_{i\alpha})p(\psi_{i\beta}) J_{i\alpha \beta}	\right)^{-1} \,,
\end{equation}
where $p(\psi_{i\alpha})$ is the relative frequency of appearance $\psi_{i\alpha}$ in the ensemble. 
The fraction $p(\psi_{i\alpha})$ is, of course, the same as $p_\alpha$ used to calculate the PaI, but note that we can ``merge'' $p(\psi_{i\alpha})$ and $p(\psi_{i\alpha'})$ by taking the same $p_\alpha$ value if $\psi_{i\alpha} = \psi_{i\alpha'}$ (the locally same community membership for node $i$), even if $\alpha \neq \alpha'$ (different memberships between $\alpha$ and $\alpha'$ appear for nodes other than $i$), to reduce the computational cost in practice. 
In other words, the MeI in Eq.~\eqref{eq:MeI} value $\Psi_{i}$ encapsulates the nodal property in terms of community membership by tracking the comembership of node $i$ across different community configurations. 
In essence, the MeI corresponds to the effective number of independent memberships of $i$ in the community ensemble, similar to that in the community level for the PaI.
 
Note that we have used another form of inconsistency measure introduced in our previous works~\cite{CoI_NJP,CoI_PRE} called \emph{companionship inconsistency} (CoI). 
We have also developed CoI as a practical tool to quantify the (in)consistency in community detection. 
Rewriting the original formula of CoI with the formulation in this paper about the community ensemble, we can express the CoI based on the cooccurrence probability between nodes $i$ and $j$, namely,
\begin{equation}\label{eq:co-occur}
W_{ij} = \sum_{\alpha=1}^{\mathcal{C}} p_{\alpha} \delta(g_{i\alpha},g_{j\alpha}) \,,
\end{equation}
which is the probability that nodes $i$ and $j$ belong to the same community (extracted by the Kronecker delta for the community labels) across different community configurations. 
For node $i$, if the values of $W_{ij}$ are $0$ or $1$ for all $j$, it is intuitive that node $i$ has consistent comembership relation with the other nodes in the network.
In contrast, if $W_{ij}$ takes intermediate values somewhere between $0$ and $1$ according to various $j$, the node $i$ is considered to have rather inconsistent relationships with the other nodes.
By symmetrizing the cooccurrence probability by squaring and taking the average, the CoI is defined~\cite{CoI_NJP,CoI_PRE} as
\begin{equation}\label{eq:CoI}
\Phi_{i} =  1 - \frac{1}{m-1} \sum_{j(\ne i)} (1 - 2W_{ij})^2 \,.
\end{equation}
The minimum level of inconsistency $\Phi_{i}=0$ is reached when $W_{ij}=0$ or $1$ for all of the other nodes $j$, and the maximum inconsistency $\Phi_{i}=1$ is reached when $W_{ij}=1/2$ for all of the other nodes $j$. 

One may wonder why we have newly developed the MeI in this paper, even if we already introduced the CoI in the previous works. 
We will answer this question by discussing the difference between the two measures in Sec.~\ref{sec:MeI_CoI_in_model_net}, the comparison of which in turn provides valuable insights about distinct characteristics of these nodal inconsistency measures in detecting the community boundary.

\section{Results}
\label{sec:results}

In this section, we present the properties of our inconsistency measures introduced in Sec.~\ref{sec:method} for various cases, and demonstrate the practicality of the inconsistency landscape in the resolution space to suggest the most statistically relevant community structures. 
Since the exact ground-truth community structures of real networks are usually unknown, we begin with the application of our measures to a series of model networks synthesized to exhibit \emph{a priori} community organizations with multiple orders of hierarchy to validate our framework. 
Along with the verification of our method to extract the most appropriate scales of communities, we also provide a microscopic property in the formation of communities from local measures of inconsistency. 
Based on the properties of our measures learned from the model networks, we turn our attention to carefully chosen real networks with different community organizations and demonstrate the ability of our method to extract valuable information on the mesoscale organizations across different scales.

\subsection{Model networks}
\label{sec:model_networks}

\subsubsection{The single-level community model network}
\label{sec:single_level_model}

\begin{figure}
\includegraphics[width=1.0\linewidth]{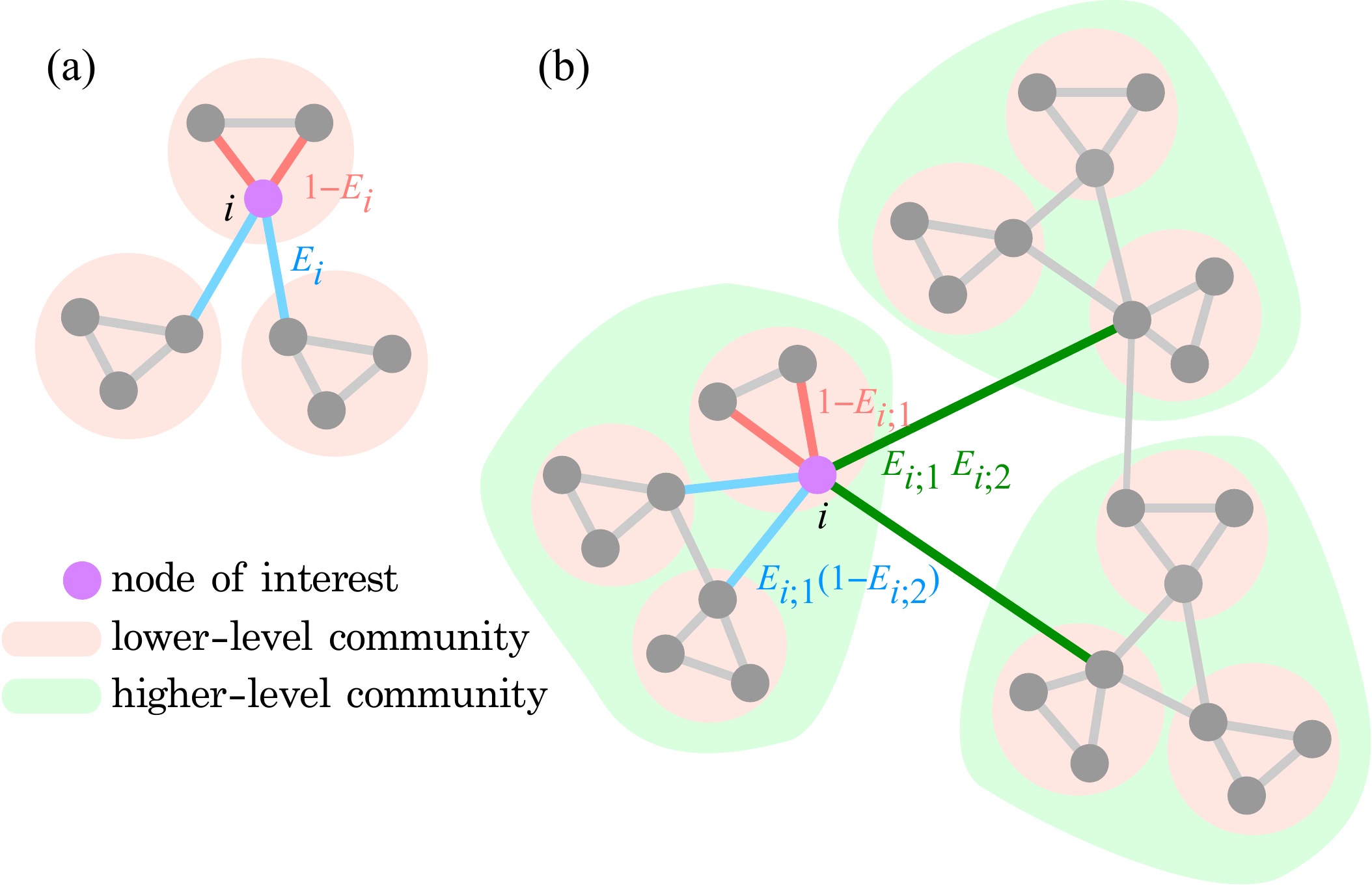}
\caption{Illustration of our model networks with different levels of prescribed communities: we use single-level (a) and double-level (b) cases as benchmark networks. In the single-level community model network, for each edge construction, a randomly selected node $i$ is connected to a node in another community with the probability equal to its externality value $E_i$ and node $i$ is connected to another node in the same community with the complementary probability $1 - E_i$. In the double-level hierarchical community model network, each node has two different externality values corresponding to the connecting probabilities to a node in different community levels.}
\label{fig:model_scheme}
\end{figure}

We first check the characteristics of several inconsistency measures for model networks with different levels of prescribed community structures, which are illustrated in Fig.~\ref{fig:model_scheme}. 
Our first synthetic benchmark network, depicted in Fig.~\ref{fig:model_scheme}(a), consists of $1200$ nodes that are equally divided into an arbitrary number $l$ of communities. 
For example, if we assign two communities in this network, $600$ nodes are allocated to each community. 
We set the prior externality value $E_i$, which is the fraction of intercommunity edges (the edges connecting the nodes belonging to different communities) of node $i$'.
Since a node with a too large value of externality tends to have more edges to another community than its own (that would violate the notion of the network community in the first place), we need to set the maximum value of prior externality as $E_{\max}$ to guarantee the existence of desired community structures. 
We assign each $E_i$ drawn from the uniform probability distribution with the range $E_i \in [0,E_{\max}]$. 
It is desirable to satisfy $E_{\max} \le (l-1)/l$, as the baseline probability of each edge belonging to other communities when we assume the absence of any community structure is $(l-1)/l$ (just the fraction of other community members) and the externality value should be less than that for a proper community structure. 
In our model, we take the convention $E_{\max} = (l-1)/l$ for simplicity. 

After dividing the nodes into preassigned communities, we create edges between nodes. 
We basically create the Erd\H{o}s-R{\'e}nyi (ER) random network~\cite{ERnet} for each community with intracommunity edges as building blocks, and sparsely connect the communities with intercommunity edges. 
The overall connection probability (edge density) is set as approximately $0.24$ corresponding to the desired number of edges $173069$, and we divide the entire set of edges into intra- and intercommunity ones. 
All of the intracommunity edges connect randomly chosen node pairs belonging to a community, and intercommunity edges connect nodes belonging to different communities.
For each edge-creation process, a node denoted by $i$ is randomly selected and connected with another randomly selected node based on the externality value $E_i$. 
For example, when $E_i = 0.4$, we connect node $i$ to a randomly chosen node in a different (the same) community with the probability $0.4$ ($0.6$), respectively. 
By repeating this process until we reach the aforementioned desired number of edges, we complete the network generation process. 
After a network is generated, with a given range of the resolution parameter $\gamma$ (used in the modularity function~\cite{fast,finding}---larger values of $\gamma$ induce a larger number of smaller communities), we apply the GenLouvain~\cite{GenLouvain} algorithm to generate the community ensemble, and observe the dependency of inconsistency measures as the functions of $\gamma$.

\begin{figure}
\includegraphics[width=1.0\linewidth]{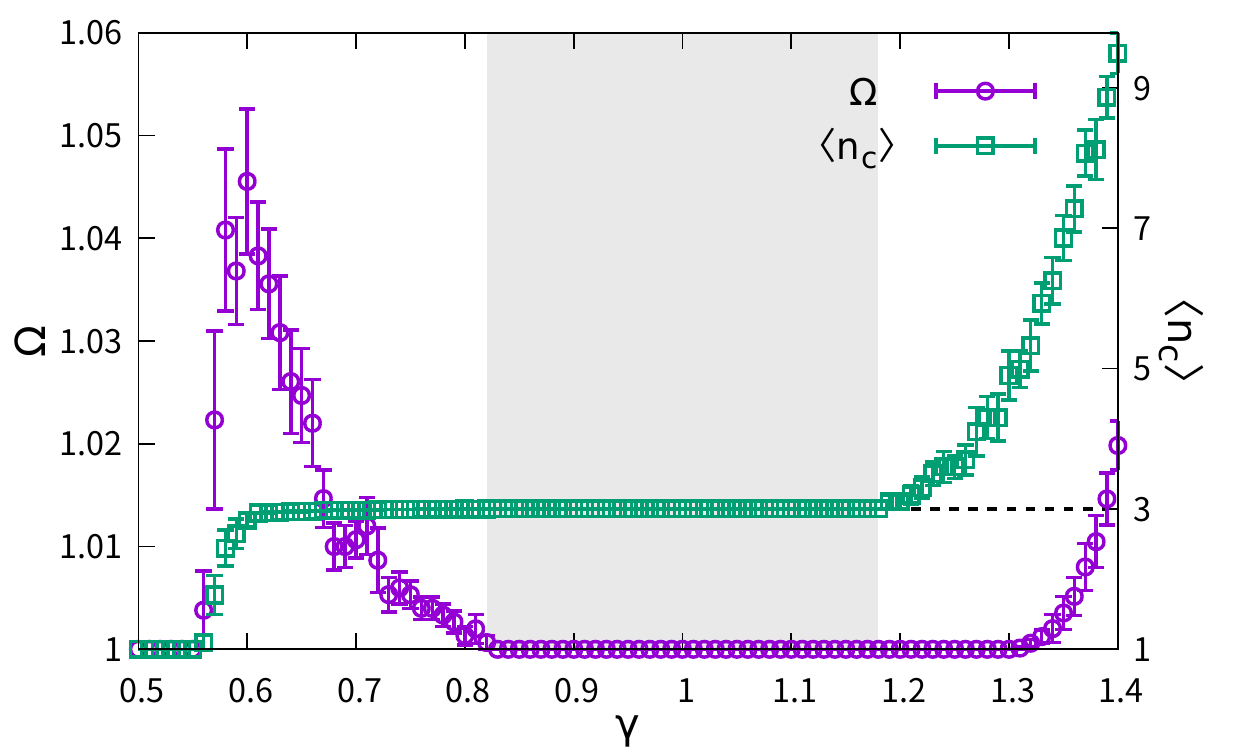}
\caption{PaI value $\Omega$ and the average number $\langle n_c \rangle$ from detected communities of the single-level community model network with three communities, as the functions of the resolution parameter $\gamma$. The error bars for both represent the standard error from network realizations. The horizontal black dashed line $\langle n_c \rangle = 3$ indicates the number of prescribed communities, which is consistent with the $\langle n_c \rangle$ values in the valid region marked by the shade.}
\label{fig:model1}
\end{figure}

In Fig.~\ref{fig:model1}, we show the result of numerical simulation for this model network with the three prescribed communities. 
We generate $10$ networks independently with the aforementioned procedure, and produce $100$ detection results for each network. 
From the results, for each network, we obtain the PaI value $\Omega$ and the average number $\langle n_c \rangle$ of communities. 
Finally, we plot the mean values with the standard errors of $\Omega$ and $\langle n_c \rangle$, averaged over the results from the $10$ networks, as we vary the resolution parameter $\gamma$ in Fig.~\ref{fig:model1}. 
We first notice that the community landscape of both quantities are roughly divided into multiple regimes. 
The values of $\Omega$ and $\langle n_c \rangle$ are flat for small values of $\gamma$, rapidly changing (a peak of $\Omega$ and rapidly increased value of $\langle n_c \rangle$) for an intermediate narrow range of $\gamma$, and they become flat again after that until they are rapidly increased. 

Let us provide phenomenological interpretation of the functional forms of $\Omega$ and $\langle n_c \rangle$, inferred from what happens during the stochastic community detection process.
First, for very small values of $\gamma$, the resolution of the detection algorithm surpasses the entire scale of the network so that all of the nodes are grouped into a single community. 
Naturally, for this trivial region, there is no inconsistency such that $\Omega = \langle n_c \rangle = 1$ by definition. 
As the resolution parameter $\gamma$ increases, the scale of detected communities begins to be compatible with the \emph{intrinsic} scales of communities in the network, so the network is divided with rather arbitrary domains first and the inconsistency starts to emerge at this ``breaking'' point. 
When the $\gamma$ value reaches appropriate scales of the communities (marked by the shaded region in Fig.~\ref{fig:model1}), the configurations inside the community ensemble become consistent and the $\Omega$ value reaches its minimum (or a small value) again. 
Most importantly, in a similar range of $\gamma$ values, the average number $\langle n_c \rangle$ of communities becomes flat with a particular integer value, three in this case as we have planted three communities before. 
In other words, this range of $\gamma$ provides the most relevant community structure with a nontrivial number ($\langle n_c \rangle > 1$) of consistent ($\Omega \approx 1$) communities. 
If we increase $\gamma$ further than that, the scale of detection starts to become incongruent with proper scales of communities again (in the extreme case of $\gamma \to \infty$, each individual node becomes its own community) and the measures become larger again.
The absence of plateau of $\langle n_c \rangle$ indicates that there is no meaningful community structure. 
In summary, the community landscape described in Fig.~\ref{fig:model1} precisely reveals the ground-truth community structure here---the three communities in a single scale.

\subsubsection{The double-level community model network}
\label{sec:double_level_model}

\begin{figure}
\includegraphics[width=1.0\linewidth]{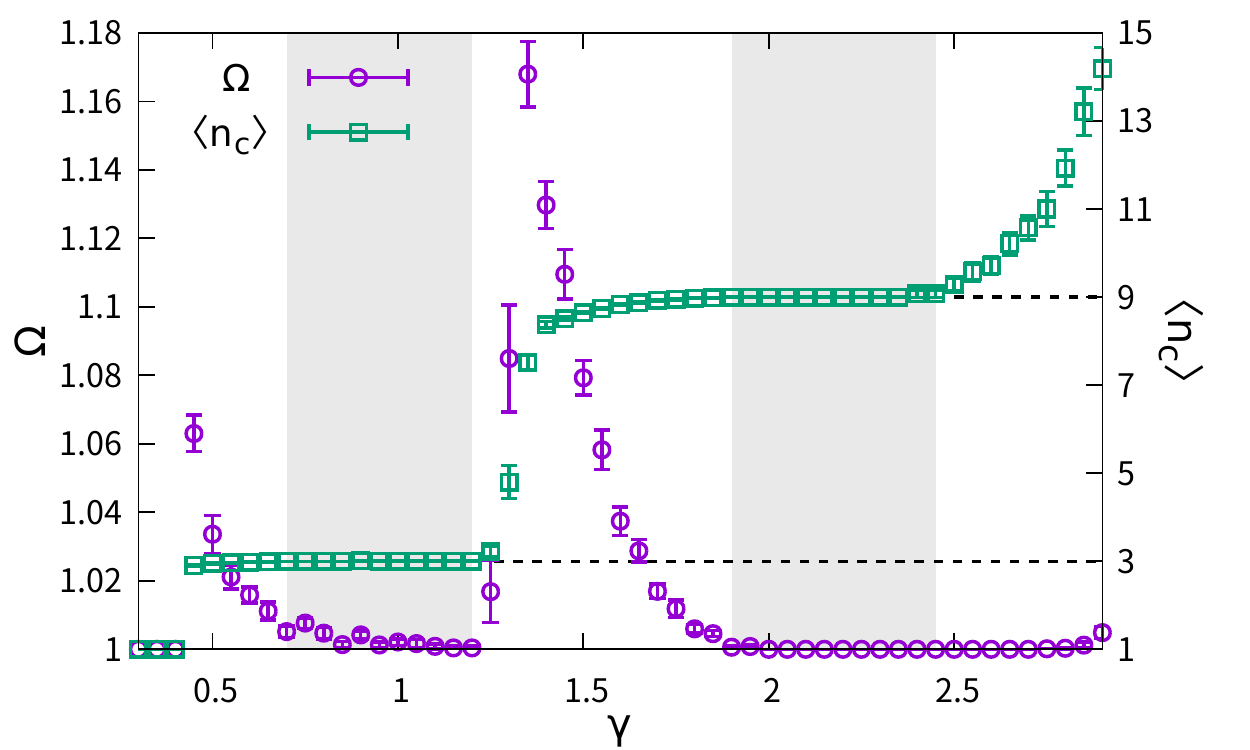}
\caption{PaI value $\Omega$ and the average number $\langle n_c \rangle$ from detected communities of the double-level community model network with three communities in the higher level and nine communities (each of the three higher-level communities includes three lower-level communities) in the lower level, as the functions of the resolution parameter $\gamma$. The error bars for both represent the standard error from network realizations. The horizontal black dashed lines $\langle n_c \rangle = 3$ and $\langle n_c \rangle = 9$ indicate the number of prescribed communities, which is consistent with the $\langle n_c \rangle$ values in the valid regions marked by the shade.} 
\label{fig:model2}
\end{figure}

\begin{figure*}
\includegraphics[width=0.9\textwidth]{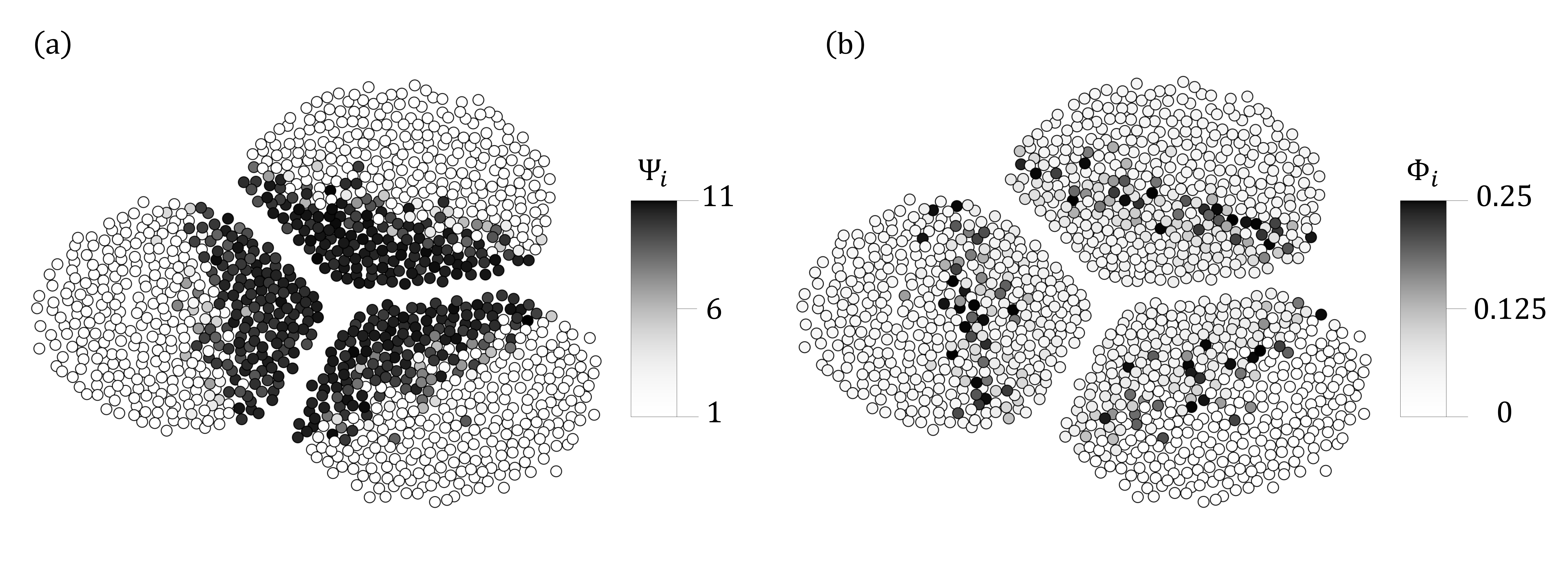}
\caption{Different nodal inconsistency measures from the same community ensemble of the single-level community model network in the case of $\gamma=1.9$. We adopt the Fruchterman-Reingold graph layout algorithm~\cite{spring} that visually separates the three prescribed communities, and omit all of the edges for simplicity. The shade of each node represents its MeI value $\Psi$ in panel (a) and the CoI value $\Phi$ in panel (b).}
\label{fig:diff}
\end{figure*}

The single-level community model composed of assembled ER random networks is understandably the most commonly used benchmark example~\cite{LFR}, but as our purpose is to extract \emph{all} of the relevant communities in \emph{various} scales, we come up with another model network. 
In our double-level community model, the assembly of communities in the small-scale level (the lower-level communities) composes the higher-level communities in a hierarchical fashion, as shown in Fig.~\ref{fig:model_scheme}(b). 
Thus the edges in this network are classified into three categories: intracommunity edges in the first level, intracommunity edges in the second level (they are intercommunity edges in the first level), and intercommunity edges in the second level (also intercommunity edges in the first level). 
As there are different levels of communities, we need two distinct externality values of each node $i$, denoted by $E_{i;1}$ and $E_{i;2}$ that correspond to the externality values with respect to the lower- and higher-level communities, respectively. 

In the edge-creation process of this network, a randomly selected node $i$ is connected to another randomly chosen node belonging to the same lower-level community as node $i$ with the probability $1-E_{i;1}$, a node belonging to the same higher-level community (but belonging to a different lower-level community) as node $i$ with the probability $E_{i;1}(1-E_{i;2})$, or a node belonging to a different higher-level (and lower-level as well by definition) community with the rest of the probability $E_{i;1}E_{i;2}$. 
In other words, each edge attached to a node provides two chances for the node to expand its interaction boundary to outside the node's own community in different levels.
In order to equalize the proportion of each category of edge, we set $E_{i;2} = 1 - E_{i;1}$ and take $E_{i;1}$ from the uniform distribution $[1/3,2/3]$ centered around the mean value $1/2$. 

In Fig.~\ref{fig:model2}, we display the inconsistency landscape of the double-level community model network. 
The network consists of $900$ nodes and $60682$ edges, which corresponds to the edge density $p = 0.15$, and three second-level communities (each of which includes $300$ nodes) in which each of them includes three first-level communities (each of which includes $100$ nodes), in the hierarchically nested way. 
Note that we use $10$ independent networks and $100$ detection results for each network, as done in the case of the single-level community model. 

Akin to the result of single-level community networks, the $\Omega$ and $\langle n_c \rangle$ values experience two abrupt changes and three stable regions. 
Note that $n_c$ values in stable areas correspond to the expected numbers of planted communities with the corresponding scales, which are $1$ (the network itself), $3$ (the higher level), and $9$ (the lower level). 
Therefore, we have demonstrated that our method successfully finds the most relevant community structures based on the comprehensive analysis of community inconsistency in various scales. 
In other words, we reveal all of the communities in different levels we have included in this hierarchically constructed model network.

\subsubsection{Local inconsistency measures in the model networks}
\label{sec:MeI_CoI_in_model_net}

The model networks introduced here are also nice test beds to illustrate the difference between the locally defined nodal inconsistency measures: MeI in Eq.~\eqref{eq:MeI} and CoI in Eq.~\eqref{eq:CoI}. 
One may imagine that the two measures describe the same type of nodal property in regard to community membership. 
However, we show that there are fundamentally different aspects between MeI and CoI, by providing the striking difference between them observed in the single-level community model in Fig.~\ref{fig:model_scheme}(a). 
In the result of the three-community model network with $\gamma = 1.9$ (Fig.~\ref{fig:diff}), which we intentionally take to highlight the situation of unstable communities (basically it tries to find spurious communities of finer scales than the ones we planted, as one can infer from Fig.~\ref{fig:model1}), we show the MeI and CoI values of each node with different shades of gray. 
In particular, pay attention to the nodes located at the central part of each panel as a result of the Fruchterman-Reingold graph layout~\cite{spring} (imposing repulsive forces between all of the node pairs and attractive forces between the connected nodes), which have inconsistent community memberships characterized by their large MeI values. 

In this case, the difference between MeI and CoI is prominent. 
The large MeI values in Fig.~\ref{fig:diff}(a) in the central region are intuitively comprehensible from the property of MeI; the nodes belong to many different small-sized communities emerged at $\gamma > 1.2$ (see Fig.~\ref{fig:model1}) that induce quite inconsistent memberships for them. 
On the other hand, as shown in Fig.~\ref{fig:diff}(b), the nodes in the most central region no longer have the largest CoI values. 
Instead, the nodes at the boundary of the center region, where $\Psi_i$ values start to decrease in Fig.~\ref{fig:diff}(a), have larger CoI values than the most central part. 

The difference comes from the definition of CoI in Eq.~\eqref{eq:CoI}, which basically measures how far each of cooccurrence probability $W_{ij}$ in Eq.~\eqref{eq:co-occur} is from $1/2$. 
In the example network here, nodes in the most central part belong to more than two distinct communities, which increases the value of CoI compared with the value of seemingly ``more internal'' nodes [dark nodes in Fig.~\ref{fig:diff}(b)] for each prescribed community. 
We have already discussed this issue of CoI in Ref.~\cite{CoI_PRE}, and from this example here we present MeI as a complementary measure to CoI in such a case. 
Simply put, the MeI and CoI measures detect the ``bulk'' and ``boundary'' of inconsistent regions, respectively. 

\subsection{Real networks}
\label{sec:real_networks}

In this subsection, we use several different types of real networks in order to investigate the detailed properties of our inconsistency measures. 
Even if the ground-truth community structure of most real-world networks is unknown, the observation of our inconsistency curve can effectively diagnose the reliability of detected communities in various scales to infer the most relevant community structures. 

\subsubsection{Zachary's karate club network}
\label{sec:karate}

\begin{figure}
\includegraphics[width=1.0\linewidth]{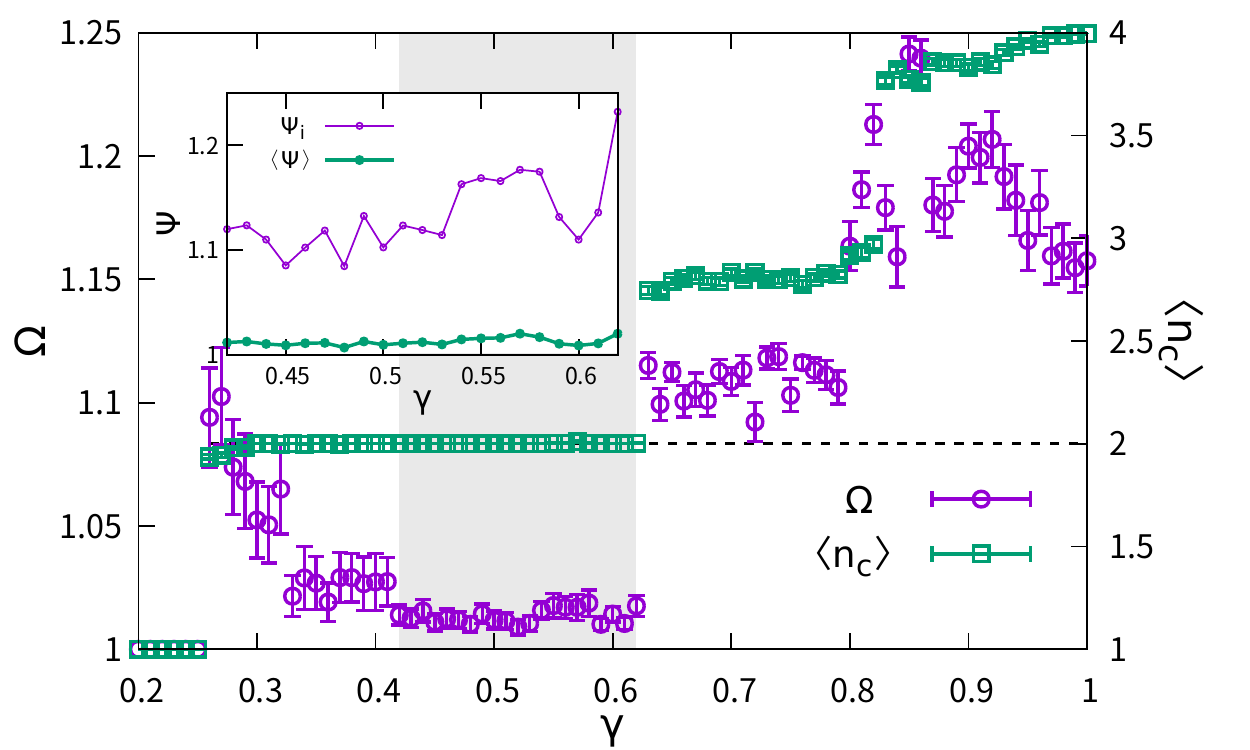}
\caption{PaI value $\Omega$ and the average number $\langle n_c \rangle$ from detected communities of Zachary's karate club network~\cite{KarateClub}, as the functions of the resolution parameter $\gamma$. The error bars of $\langle n_c \rangle$ and $\Omega$ are the standard errors calculated from $500$ independent community detection results and the randomly partitioned $50$ equal-sized subsamples, respectively. The most valid range of $\gamma \in [0.42,0.62]$ based on the lowest values of $\Omega$ with small errors is shaded, and the horizontal black dashed line $\langle n_c \rangle = 2$ indicates the suggested number of communities in the valid region. The inset shows the MeI value $\Psi_i$ of the most inconsistent node $i$ (see Fig.~\ref{fig:karate_0.5}) in the valid region of $\gamma$, compared with the MeI value $\langle \Psi \rangle$ averaged over all of the nodes.}
\label{fig:karate}
\end{figure}

\begin{figure}
\includegraphics[width=1.0\linewidth]{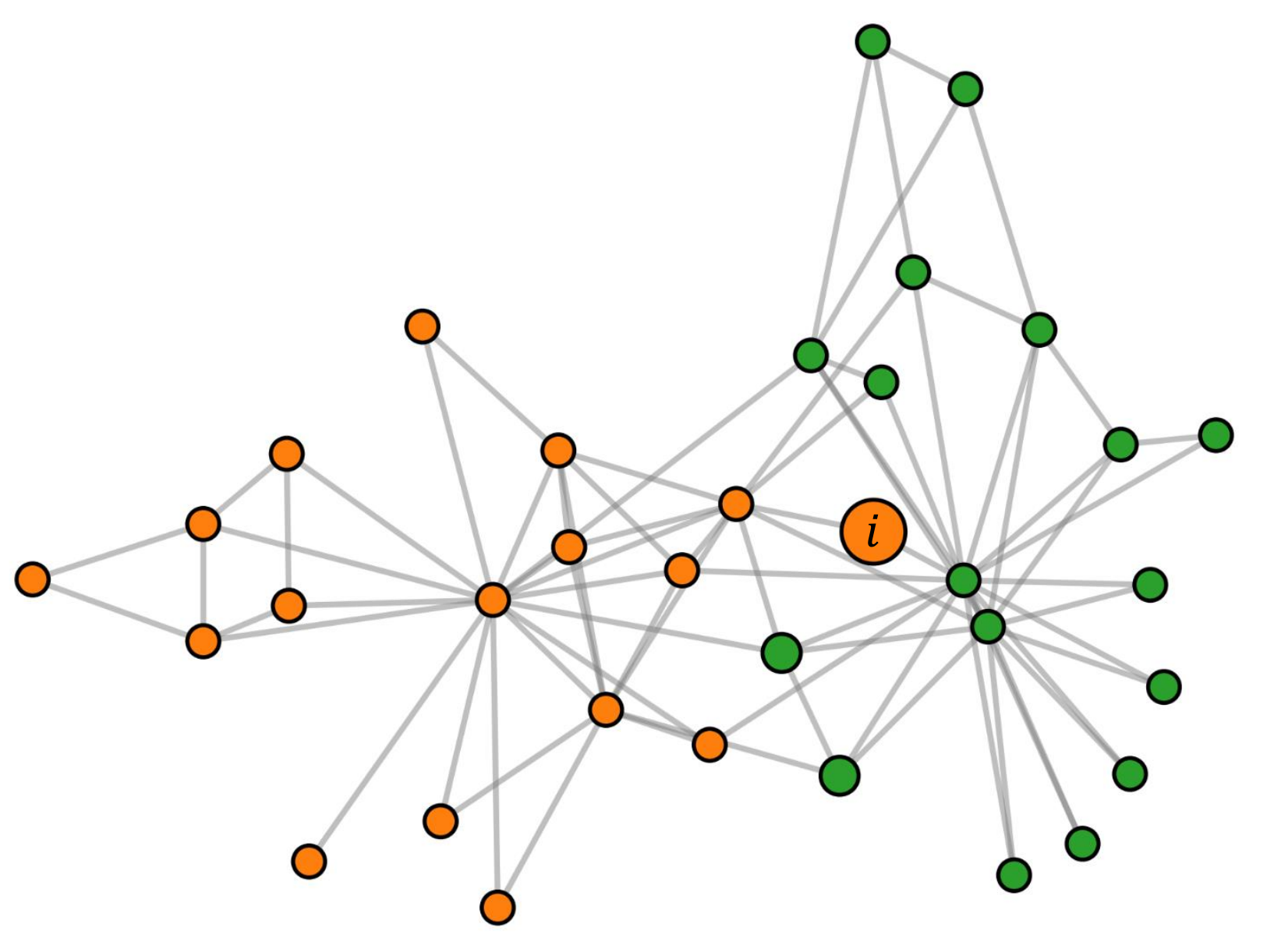}
\caption{Community structure of the karate club network~\cite{KarateClub} at $\gamma=0.5$. The color of each node represents its community identity from the most probable configuration, and we set the size of each node proportional to $\sqrt{\Psi-1}$ for better visibility. We indicate the most inconsistent node with the maximum $\Psi$ value as $i$, which maintains the largest $\Psi$ in the valid range $[0.42,0.62]$ (see the inset of Fig.~\ref{fig:karate}).}
\label{fig:karate_0.5}
\end{figure}

We begin our analysis with the most well-known network used as a benchmark in the field of community detection: Zachary's karate club network that represents the relations between trainees in a university karate club~\cite{KarateClub}. 
It is known that the network consists of two separate communities, triggered by the conflict between the administrator and the master. 
We plot the PaI value $\Omega$ along with the average number $\langle n_c \rangle$ of communities as the functions of the resolution parameter $\gamma$ in Fig.~\ref{fig:karate}. 
The result is from a community ensemble formed by $500$ independent community detection results, which we believe constitute a statistically valid ensemble judged by the small values of standard error from resampling processes described in Fig.~\ref{fig:karate}. 
Compared with the results of the model networks in Sec.~\ref{sec:model_networks}, the $\Omega$ and the $\langle n_c \rangle$ values as the functions of $\gamma$ show a more diverse pattern. 

Observing the $\langle n_c \rangle$ curve, we find four flat ranges corresponding to one, two, three, and four communities, respectively. 
Therefore, one may conclude that all of the four regimes represent valid community structures judged only by the number of communities. 
However, this network provides a perfect example of the importance of taking the inconsistency of communities into account. 
Besides the trivial regime of $\langle n_c \rangle = 1$, the PaI value $\Omega$ rapidly increases around $\gamma \approx 0.26$ first and decreases again around $\gamma \approx 0.42$, which corresponds to the $\gamma$ regime that starts to detect two communities, as for the ground-truth communities in the model networks shown in Fig.~\ref{fig:model1} (three communities) and Fig.~\ref{fig:model2} (three and nine communities). 
For the resolution parameter range $\gamma = [0.42, 0.62]$, the system shows the signature of reliable and consistent community structure characterized by a stable integer value of $\langle n_c \rangle = 2$ on top of small $\Omega$ values with small standard errors. 
For $\gamma > 0.62$, the $\langle n_c \rangle$ curve shows rather flat regimes around $\langle n_c \rangle \approx 3$ and $\langle n_c \rangle \approx 4$, but $\Omega$ never returns to a small value as in the range $\gamma = [0.42, 0.62]$. 
In other words, even if the algorithm detects some smaller-scale communities than the two communities, as in the case of model networks, those smaller-scale communities are not concomitant with the intrinsic mesoscale structure of the karate club network. 
As a result, we propose that the range $\gamma = [0.42, 0.62]$ where two communities are detected is the most valid resolution and the community structure of this network, confirmed by the original work on this network~\cite{KarateClub} that assumes the two conflicting communities.

Figure~\ref{fig:karate_0.5} shows the community structure and the local inconsistency represented by the MeI value $\Psi$ for each node of the karate club network, detected with the resolution parameter $\gamma = 0.5$ in the valid regime. 
The color of each node denotes the community identity of the most probable configuration, and the node size represents the relative $\Psi$ value in community ensemble. 
We check that most nodes display consistent memberships with small $\Psi$ values (see the inset of Fig.~\ref{fig:karate}), except for a single node located at the boundary between two communities, as also demonstrated by the large CoI value in Ref.~\cite{CoI_PRE} and discussed in the original work~\cite{KarateClub}. 
The local inconsistency of a specific node denoted by $i$ in Fig.~\ref{fig:karate_0.5} maintains in the entire consistent region $\gamma= [0.42, 0.62]$, as shown in the inset of Fig~\ref{fig:karate}. 
Therefore, by observing the local inconsistency measure on top of the global measure in various scales, we can investigate detailed community structures in a comprehensive way; in this case, first we extract the most valid community structure and pinpoint locally consistent and inconsistent nodes in the community structure we can trust.

\subsubsection{The Chilean power-grid network}
\label{sec:power_grid}

\begin{figure}
\includegraphics[width=1.0\linewidth]{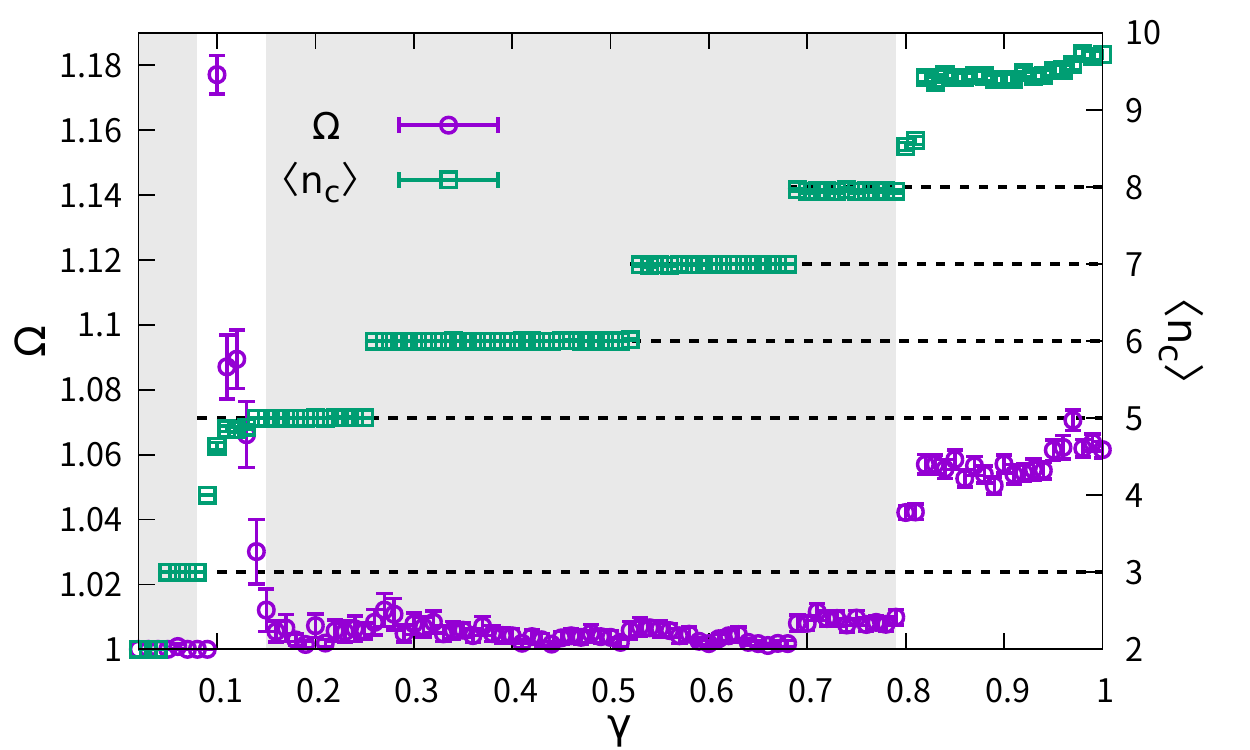}
\caption{PaI value $\Omega$ and the average number $\langle n_c \rangle$ from detected communities of the Chilean power-grid network~\cite{CL_data}, as the functions of the resolution parameter $\gamma$. The error bars of $\langle n_c \rangle$ and $\Omega$ are the standard errors calculated from $500$ independent community detection results and the randomly partitioned $50$ equal-sized subsamples, respectively. The most valid ranges of $\gamma$ based on $\Omega$ and $\langle n_c \rangle$ are shaded, and the horizontal black dashed lines $\langle n_c \rangle = 3, 5, 6, 7$, and $8$ indicate the suggested numbers of communities in the valid regions.}
\label{fig:Rgrid}
\end{figure}

The Chilean power-grid network represents the interconnected topology of large infrastructures that maintain the electrical power system of the central region of Chile. 
Each node in this network represents an electric power system facility such as power plants and substations, and edges correspond to the high voltage transmission lines between the nodes. 
We use the ``without-tap'' (WOT) version of the Chilean power grid network presented in Ref.~\cite{CL_data} (see the data description therein for details), which is considered as an effective simplification of a high-voltage power transmission structure. 
Since the network topology of a power-grid system literally forms the backbone behind its stability and efficiency, our consistency analysis applied to the network provides the first clue to pinpoint the relevant scales of functional units to maintain its stable operation.

The result of consistency analysis on the Chilean power-grid network shows several interesting features. 
From the result reported in Fig.~\ref{fig:Rgrid}, we first notice that the $\langle n_c \rangle$ curve shows several steps of plateaus in a wide range of the resolution parameter $\gamma$. 
This type of intermittent increment of $\langle n_c \rangle$ is observed in the case of the karate club (Fig.~\ref{fig:karate}) as well, but there exists a crucial difference in the inconsistency indicated by $\Omega$. 
Contrary to the karate club network, the PaI value $\Omega$ stays as quite small values across different numbers of communities, i.e., for $\langle n_c \rangle = 3, 5, 6, 7$, and $8$ over a wide range of $\gamma$. 
A natural explanation is the existence of multi-level communities as demonstrated in the hierarchically constructed double-level community model in Fig.~\ref{fig:model2}.
Therefore, we conclude that the Chilean power-grid network is also composed of multi-level communities, possibly in a hierarchical fashion. 
One particularly peculiar feature is the fact that although $\langle n_c \rangle = 3$ (the leftmost shaded region in Fig.~\ref{fig:Rgrid}) and $\langle n_c \rangle = 5$ are separated by the sharp peak of $\Omega$ around $\gamma \approx 0.12$, the stable regime of $\Omega \approx 1$ spans the range of multiple integer values of $\langle n_c \rangle$ without interruption. 
It indicates the absence of unstable scales of communities within the corresponding range (the second shaded region that takes the majority of the $\gamma$ range in Fig.~\ref{fig:Rgrid}) and it is worthwhile to investigate the $\gamma$ range further by taking the local inconsistency measure.

\begin{figure*}
\includegraphics[width=0.7\textwidth]{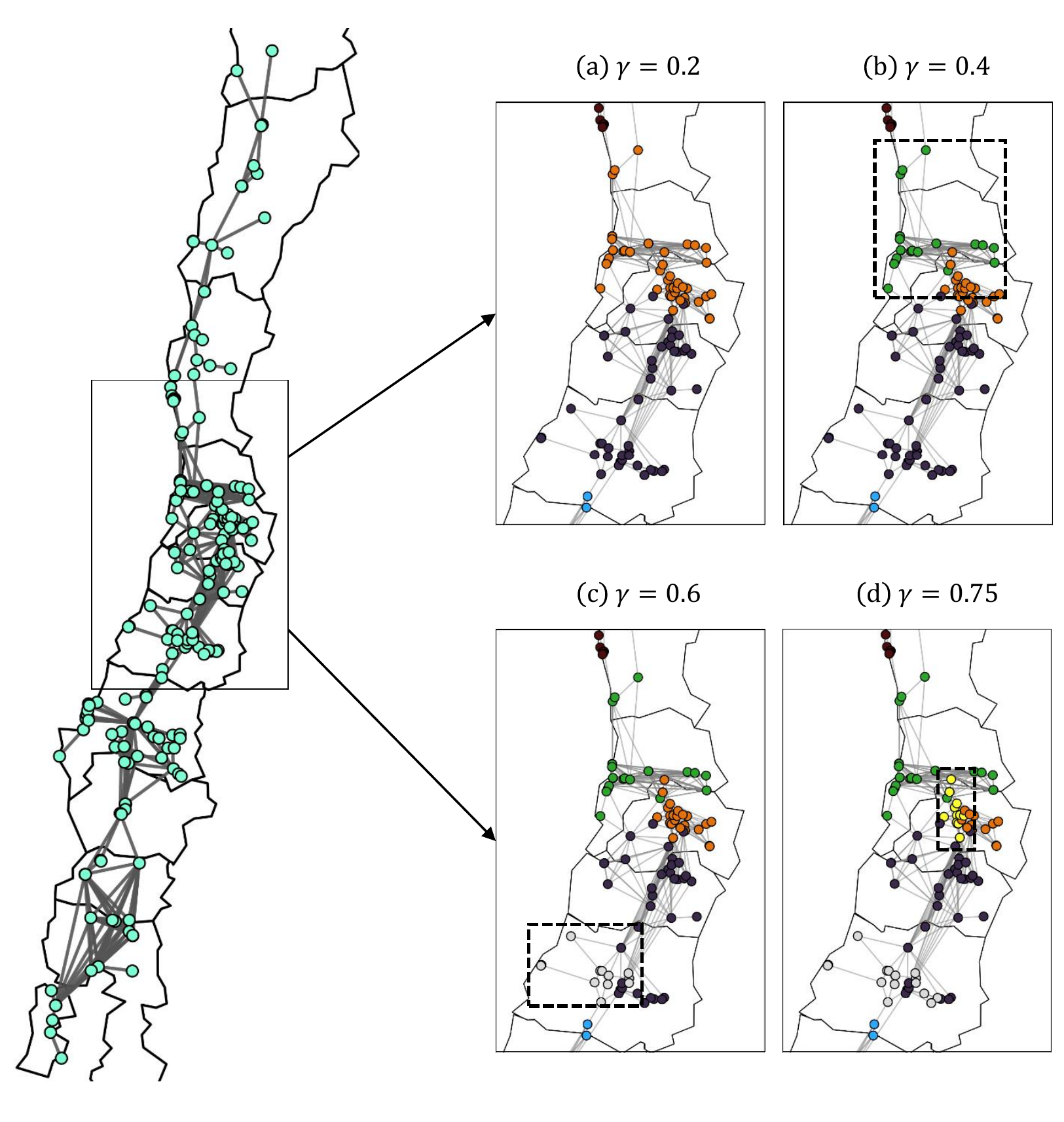}
\caption{Examples of the community structures of the Chilean power-grid network for different values of $\gamma$. In each plot, 
the color represents the community identity of the nodes in the most probable community configuration in the community ensemble.  
We focus on the central region of Chile as a representative example.}
\label{fig:Rgrid_dist}
\end{figure*}

In order to visually inspect the detected community structures including the local inconsistency measure for this particular $\gamma$ range, we show the community structures with the MeI values $\Psi$ on the nodes for several $\gamma$ values in Fig.~\ref{fig:Rgrid_dist}. 
We choose four different $\gamma$ values, which correspond to $\langle n_c \rangle = 5, 6, 7$ and $8$, respectively. 
As we increase $\gamma$ values, it is observed that new communities enclosed in existing communities emerge. For example, the orange community for $\gamma=0.2$ in Fig.~\ref{fig:Rgrid_dist}(a) splits into the orange and green communities for $\gamma=0.4$ in Fig.~\ref{fig:Rgrid_dist}(b). 
The black community in the same panel is divided into the black and white communities for $\gamma=0.6$ in Fig.~\ref{fig:Rgrid_dist}(c). 
Finally, the orange community in Fig.~\ref{fig:Rgrid_dist}(c) is further divided into the orange and yellow communities for $\gamma=0.75$ in Fig.~\ref{fig:Rgrid_dist}(d).
All of the newly emerged community structures in this example are completely included in existing communities, which is the definition of the hierarchical community structure. 
In fact, we have speculated on such a hierarchical structure based on the distribution of CoI values for a specific $\gamma$ value in Ref.~\cite{CoI_PRE}, and it is finally confirmed with the inconsistency landscape analysis in this paper. 
Those community structures spanning different scales provide valuable information applicable for maintaining and planning the power transmission systems, in particular, when various scales of spatial environments in regard to renewable energy sources become more important in the future.

\subsubsection{The photosynthesis-system network}
\label{sec:PS2}

\begin{figure}
\includegraphics[width=1.0\linewidth]{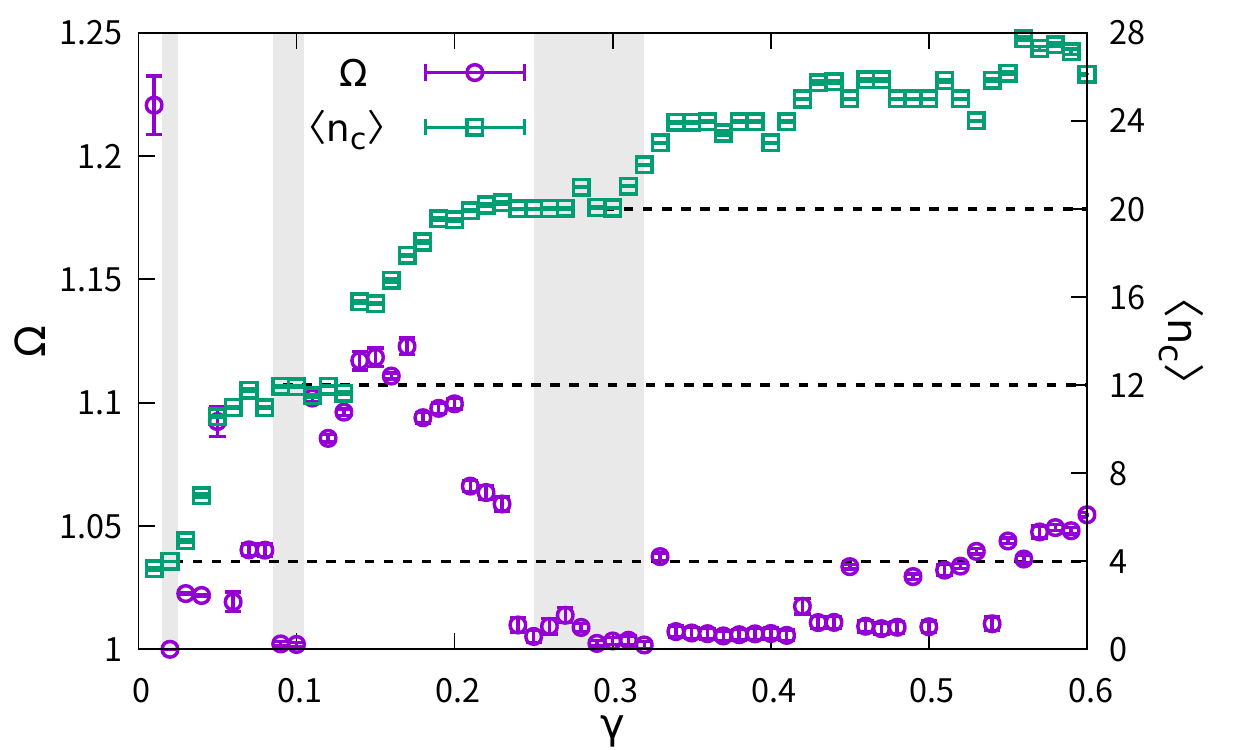}
\caption{PaI value $\Omega$ and the average number $\langle n_c \rangle$ from detected communities of the PS2 network~\cite{HKim2021}, as the functions of the resolution parameter $\gamma$. The error bars of $\langle n_c \rangle$ and $\Omega$ are the standard errors calculated from $500$ independent community detection results and the randomly partitioned $50$ equal-sized subsamples, respectively. The most valid ranges of $\gamma$ based on $\Omega$ and $\langle n_c \rangle$ are shaded, and the horizontal black dashed lines $\langle n_c \rangle = 4, 12$, and $20$ indicate the suggested numbers of communities in the valid regions.}
\label{fig:PS2}
\end{figure}

\begin{figure*}
\includegraphics[width=0.8\textwidth]{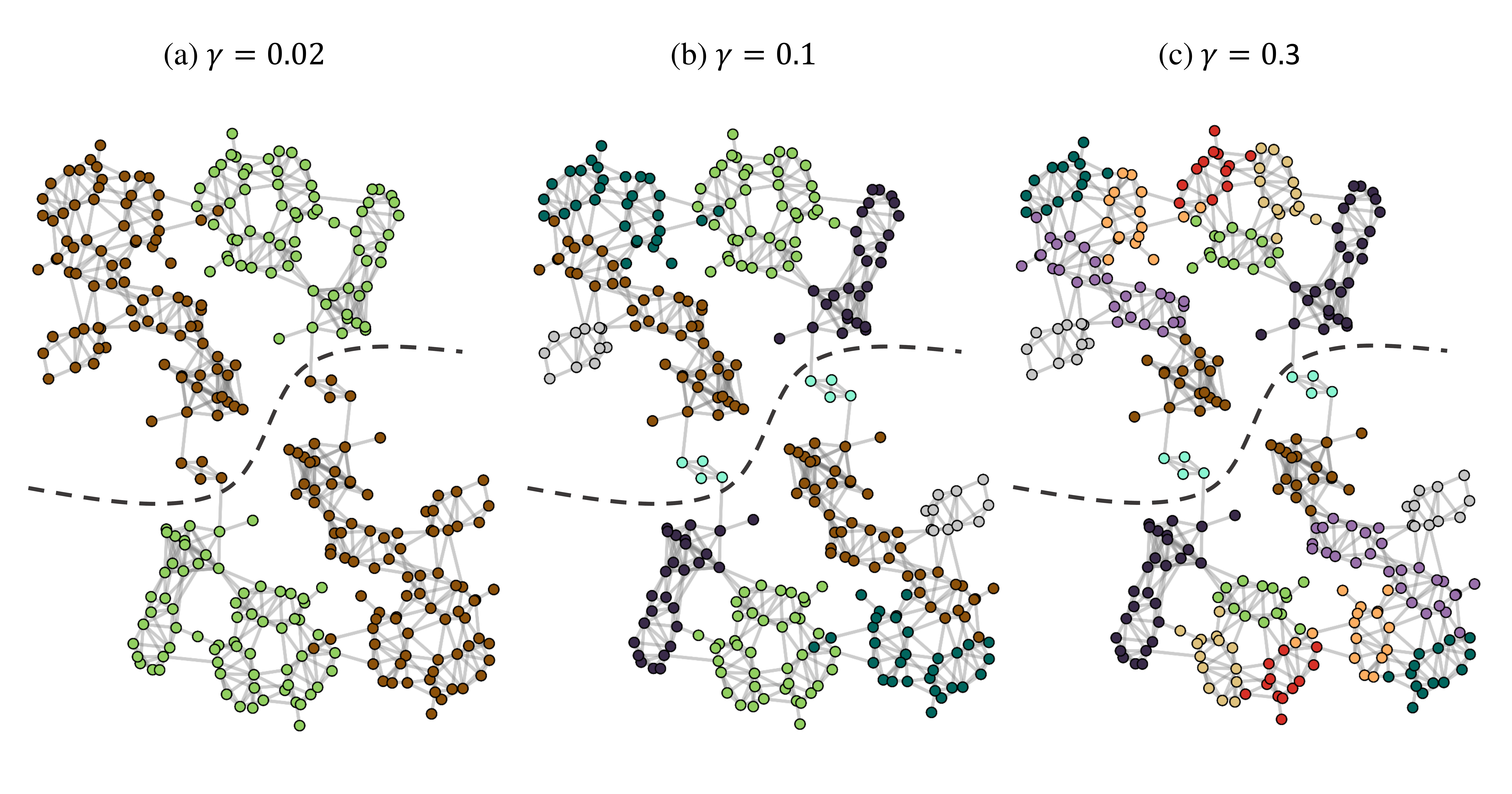}
\caption{Illustrative examples of community structures of the PS II network. We only show the edges with the weights larger than $0.4$. Since the network has the point symmetry with respect to the center, we plot the black dashed curve that separates the different sides, and assign the same color to each community pair located in the opposite direction (shown as the mirror image) for better community identification.}
\label{fig:PS2_dist}
\end{figure*}

We move from the most important macroscale energy transmission system of the humankind to a nanoscale energy transfer system in nature. 
The photosystem II supercomplex (PS2) network in plant cells describes the interactions between biological units that play a major role during photosynthesis by absorbing energy from light and transporting it to the reaction centers in the system. 
The PS2 network consists of various proteins composed by multiple chlorophylls that capture photons and then transfer the photons between them. 
In the PS2 network, each node represents a chlorophyll and a directed edge between a pair of nodes represents the energy transition rate that is calculated by the distance and the difference in the molecular orientation between the nodes~\cite{Renger:2009dn,Mazor:2017hi,HKim2021}. 
To analyze the network within the current framework of community detection,\footnote{The community detection in directed networks is certainly an interesting topic, but it is rather nontrivial, from the very definition of community itself. See Refs.~\cite{Leicht2008,YKim2010} for details.} we first symmetrize the bilateral weights by assigning the undirected weight between nodes $i$ and $j$ as $w_{ij} \equiv (w_{i \to j} + w_{j \to i})/2$, where $w_{i \to j}$ is the directed weight from node $i$ to node $j$ in the original network. 
Then, the symmetric weight $w_{ij}$ between nodes $i$ and $j$ represents the relatedness between them in the photosynthetic process. 
As the weighted edges are always defined from their three-dimensional spatial structure for the transition rate, the PS2 network is fully connected but most of the significant edges are between close pairs as the weight is proportional to $d^{-6}$, where $d$ is the distance between the magnesium atoms located in the central part of a node. 
The GenLouvain algorithm~\cite{GenLouvain} can fully take advantage of the weight information, which we use to analyze the weighted PS2 network.

The inconsistency plot shown in Fig.~\ref{fig:PS2} of the PS2 network reveals a more complicated structure to interpret than the other examples presented so far. 
The inconsistency landscape is basically composed of narrow consistent regions separated by wide inconsistent regions. 
The consistent regions where $\Omega \approx 1$ are around $\gamma = 0.02, 0.1$, and $0.3$. 
We illustrate the actual community structures for those $\gamma$ values in Fig.~\ref{fig:PS2_dist}, which also exhibits hierarchical community structures as in the Chilean power-grid network (Fig.~\ref{fig:Rgrid_dist}). 
It should be noted that the spatial plot of the network displays the point symmetry with respect to the center, so that the community structure is also exactly symmetric along the entire resolution. 
Most notably, one can see the hierarchically nested community structures when we vary the resolution parameter $\gamma$ here as well, where large communities for small $\gamma$ values are divided to smaller communities for larger $\gamma$ values. 
Since most of the community structures adequately describe the geographical closeness, pinpointing the valid resolutions for network communities can provide insightful clues to elucidate detailed spatial structures in biological systems~\cite{SHLee2019}. 

\section{Summary and Discussion}

In this paper, we have presented a practical methodology to quantify and utilize the global and local consistency of community structures generated by stochastic community detection. 
We argue that our approach on network communities is useful not only to identify particular community structures by providing the guideline to find the most reliable communities, but also to investigate from the most global information of the overall community landscape to the most local information on nodal inconsistency. 
With the resolution landscapes of the network models with single and double levels of communities, we have confirmed that our method successfully extracts the predefined ground-truth mesoscale substructures of different scales, in particular, even when the model network includes a hierarchically nested community structure.
Based on the properties of inconsistency measures discovered for the model networks, we have applied our method to representative real networks and found a number of meaningful structures across different scales. 
The karate club network demonstrates the role of a locally inconsistent node between communities, and the Chilean power-grid and the photosynthesis-system networks reveal the hierarchically organized community structures of various types: stable communities existing across wide or narrow ranges of resolution.

In the general context of community detection, we can interpret the concept of community consistency as a useful toolkit to take an overall view of community landscape in terms of statistical reliability. 
The original purpose of stochastic detection algorithms is to find ``the most appropriate'' community structure, but, by their nature of stochasticity, we can estimate the selection probability of different configurations depending on the intrinsic mesoscale structures of networks. 
Note that the probability of each candidate configuration effectively corresponds to the basin of attraction in the (high-dimensional) community-configuration space, of which we take advantage to clarify the reliability of each configuration. 
In this spirit, generating a community ensemble is equivalent to exploring the configuration space, where the size of the ensemble and the accuracy in the sampling scheme could be important. 
A systematic study on such a practical issue of sampling can be a good candidate for future work. On the local inconsistency, we have compared the MeI measure with the CoI measure we introduced before, which provide an insightful difference between their viewpoints on community boundaries. 
In summary, the concept of community consistency is more than just about finding communities; it is an effective set formed by fundamental aspects of network topology.

Some particular subsets of the consistency landscape can also be worth investigating further. For instance, Ref.~\cite{convex} exploits the relations between optimal configurations across different scales. By extrapolating the modularity values corresponding to the optimal configurations in terms of the convex hull problem, they provide a landscape of optimal configurations in different scales. Interestingly, their method to find the optimal values of modularity and ours to find the optimal resolution values shed light on complementary parts of community structures, so comparing them could be inspiring future work. 

As we mentioned in Sec.~\ref{sec:intro}, in a practical sense as well, the decision of the appropriate resolution of communities is crucial. 
In comparison to the previously developed \emph{ad hoc} heuristics that highly depend on properties of specific network topology, our consistency-based method established on the landscape in the resolution space provides a principled criterion to choose the most suitable community scales of networks of interest. Finally, we also would like to remark on several potential limitations of our method that are worth investigating in future work. For instance, there are arbitrary choices in the range of resolution scanned to obtain the final consistency plot, in the threshold values of $\Omega$ to pinpoint the consistent region, and, last but not least, in the types of stochastic algorithm used to detect communities. On the possible algorithm dependency, we have verified that the recently developed Leiden algorithm~\cite{Leiden} produces a qualitatively similar result to the result from the GenLouvain algorithm in this paper. 
In addition, there are some prospective topics related to the further understanding of community structure, e.g., the relationship between multiscale community structures, the conspicuous change of membership across different scales of communities, and practical interpretations and more applications of local consistency. We hope that our study can be the first guide to approaching such rich topics on this subject.

\begin{acknowledgments}
D.L. and B.J.K were supported by ``Human Resources Program in Energy Technology'' of the Korea Institute of Energy Technology Evaluation and Planning (KETEP), granted financial resource from the Ministry of Trade, Industry $\&$ Energy, Republic of Korea (Grant No. 20194010000290), S.H.L. was supported by the National Research Foundation (NRF) of Korea Grants No. NRF-2018R1C1B5083863 and No. NRF-2021R1C1C1004132, and H.K. was supported by the National Agency of Investigation and Development, ANID, through the grant FONDECYT No. 11190096.
\end{acknowledgments}

\bibliographystyle{apsrev4-2}


%

\end{CJK}
\end{document}